\documentclass[twocolumn, prl, floatfix]{revtex4}

\usepackage{graphicx}% Include figure files
\usepackage{bm}% bold math

\begin{document}

\title{Identifying Nearby UHECR Accelerators using UHE (and VHE) Photons}

\author{Taylor,~A.~M.}
\affiliation{Max-Planck-Institut f\"ur Kernphysik, P.O. Box 103980, D 69029 Heidelberg, Germany}
\author{Hinton,~J.~A.}
\affiliation{School of Physics \& Astronomy, University of Leeds, Leeds LS2 9JT, UK}
\author{Blasi,~P.}
\affiliation{INAF/Osservatorio Astrofisico di Arcetri Largo E. Fermi, 5 50125 Firenze, Italy}
\author{Ave,~M.}
\affiliation{Excellence Cluster Universe, Technische Universit\"at M\"unchen, Boltzmannstr. 2, 
D-85748, Garching, Germany}
\affiliation{University of Chicago, Enrico Fermi Institute, Chicago, IL, USA}
\date{\today}

\begin{abstract}

Ultra-high energy photons (UHE, $E>10^{19}$~eV) are inevitably produced during the propagation 
of $\sim$10$^{20}$~eV protons in extragalactic space.
Their short interaction lengths ($<$20~Mpc) at these energies,
combined with the impressive sensitivity of the Pierre Auger Observatory detector to these 
particles, makes them an ideal probe of nearby ultra-high-energy cosmic ray (UHECR) sources.
We here discuss the particular case of photons from a single nearby (within 30~Mpc) source 
in light of the possibility that such an object might be responsible for several
of the UHECR events published by the Auger collaboration.
We demonstrate that the photon signal accompanying a cluster of a few $>6\times10^{19}$~eV UHECRs 
from such a source should be detectable by Auger in the near future.
The detection of these photons would also be a signature of a light composition of the 
UHECRs from the nearby source.

\end{abstract}

\pacs{Valid PACS appear here}% PACS, the Physics and Astronomy
                             % Classification Scheme.
%\keywords{cosmic rays}%Use showkeys class option if keyword display desired

\maketitle

The origin of ultra-high-energy cosmic rays (UHECR, $E>10^{19}$~eV) still remains a mystery.
However in recent years clear progress has been made, 
with both the HiRes \cite{hiresgzk} and the Pierre Auger Observatory (PAO) \cite{Abraham:2008ru} 
providing strong evidence of a suppression in the cosmic ray spectrum above $\sim 5\times 10^{19}$~eV.
Moreover, the PAO has presented the first evidence that the arrival distribution of cosmic rays with 
energies larger than $\sim 6\times10^{19}$~eV is anisotropic \cite{AugerCollaboration:2007}.
If the flux suppression is interpreted as a consequence of severe proton energy losses during their 
propagation in extragalactic space, ie. the GZK feature \cite{Greisen:1966jv,Zatsepin:1966jv}, the sources 
of $\gtrsim 6 \times10^{19}$~eV UHECRs would be required to reside in the nearby Universe, 
within several hundred Mpc of Earth. This anisotropy could therefore be a natural consequence of 
the fact that the local universe is inhomogeneous.

Presently indirect measurements of the UHECR composition provide an unclear picture. 
Measurements of the $X_{\rm max}$ value of air showers detected by the PAO \cite{Unger:2007mc}
seem to indicate an increasingly heavy composition of UHECR whereas measurements by HiRes 
\cite{Archbold:2003vw} indicate a predominantly light composition.
Such measurements are of crucial importance in establishing whether the sources accelerate 
predominantly protons or heavier nuclei and hence in determining the expected deflection of
UHECR in both Galactic and extragalactic magnetic fields.
Though the anisotropy signal observed by the PAO is perhaps more simply explained if the 
composition is light, interpretations based on a few nearby sources of heavy nuclei 
are also possible \cite{Wibig:2007pf}. In either case propagation losses limit the UHECR 
horizon distance to below $\sim$100~Mpc at energies $\geq$10$^{20}$~eV. 
In this letter we assume that UHECRs with energies $>6\times 10^{19}$~eV are protons.

For UHECR protons, the main channel of energy losses at energies $\gtrsim 5\times 10^{19}$~eV is 
photopion production, with $\sim$0.5~mb cross-sections and inelasticities of the order $m_{\pi}/m_{p}$ 
at threshold. Following their production and escape from the source region, 
UHECR protons undergo these photo-pion interactions 
with background radiation fields (predominantly the CMB) producing both 
neutral and charged pions and hence a secondary UHE photon and neutrino flux around
the source region.
With the propagation of these UHE-photons in turn being limited through $\gamma\gamma$ interactions 
with background photons, a region of space should exist around the source in which 
this UHE-photon flux is maximal.  With the $\gamma\gamma$ cross-section peaking at a value 
of roughly 200~mb, and the dominant photon background target to these photons, the cosmic radio background 
(CRB), having a density of $\sim$1~cm$^{-3}$, the interaction lengths for this process, 
above threshold, are also $\sim$ a few Mpc. Thus, the distance for which the generated 
UHE-photon flux is maximum would lie between a few Mpc and a few tens of Mpc from the source.
% POSSIBLE ADDITION
We note here that though the CMB presents a much larger photon background target in terms
of number density, the interaction of UHE-photons (and UHE-electrons) with it is suppressed due to the
Klein Nishina suppression in the cross section for center-of-mass energies $\gg m_{e}c^{2}$.

For sources more than a few tens of Mpc away the energy in the produced UHE photon population may go into 
the development of a full electromagnetic cascade, with the energy flux arriving mainly at GeV-TeV 
energies, such sources either contributing to the diffuse gamma-ray background at these energies, 
or being individually resolvable in some scenarios \cite{blasicas,gabici}, depending on the strength
of the inter/extragalactic magnetic fields. 
However, for a more nearby source, the electromagnetic cascade's development has insufficient
time to develop, with only the appreciable production of a first generation of UHE photons,
leading to their contribution to the diffuse UHE photon background. Thus, the
search for UHE-photons in the arriving UHECR flux can provide information about the
local UHECR sources on these scales \cite{Aharonian:1992qf}.
The level of this background at 10$^{19}$~eV is already strongly bounded by PAO observations 
to be less than $\sim$2\% of the proton flux \cite{Collaboration:2009qb}.
This value exceeds by a factor between a few and a few tens that expected from theoretical 
calculations \cite{Gelmini:2005wu,Taylor:2008jz}.
However, as demonstrated by \cite{Taylor:2008jz}, an enhancement of the UHE photon flux 
can be expected in the presence of a local source.

A nearby source may easily result in the production of multiplets of UHECR events 
coming from within a few degrees on the sky, provided the composition at the source is 
dominated by protons at the highest energies to avoid sizable deflections in the 
intervening intergalactic magnetic fields (IGMFs).
Furthermore, it has been suggested \cite{Moskalenko:2008iz} that several of the 27~PAO 
events with energy $>6\times 10^{19}$~eV used in the anisotropy study 
originate from the nearest active galaxy -- Centaurus A.
At a distance of only $\sim$3.8~Mpc \cite{Rejkuba:2003xt} and with an exposure to the PAO of
1200~km$^{2}$~sr, if 2 of the PAO events originate from Cen A, the required luminosity 
to account for these events, assuming a source
spectrum with spectral index 2 between 10$^{19}$~eV and the cutoff, 
is $\sim 0.4 \times 10^{40}$~erg~s$^{-1}$. This is a small
fraction of its total bolometric luminosity \cite{Evans:2004ev} and comparable to the
inferred TeV luminosity as recently reported by the H.E.S.S. collaboration~\cite{Aharonian:2009xn}.
In this way, Cen A may be considered as a promising UHECR source candidate.  
Indeed the possible subsequent (cascade) TeV photon and UHE neutrino fluxes from Cen A have recently been 
calculated for this scenario \cite{Kachelriess:2008qx}.
However, we note that it is still not clear if any region, within an object such as Cen A (jets, inner 
and outer lobes, black hole) meets the basic criteria for UHECR acceleration 
(see e.g. \cite{Norman:1995,Casse:2001be,O'Sullivan:2009sc}).

Following the analytic description of the photon flux development in \cite{Taylor:2008jz}, 
for a very nearby ($<10$~Mpc) object such as Cen A, a simple expression can be 
used to estimate the UHE photon flux coming from it. 
For this source most of the arriving protons have yet to undergo pion production interactions
and the photon flux is dominated by the first generation of photons generated through 
$\pi^{0}$ decay.
% POSSIBLE ADDITION
Through a consideration of the two coupled differential equations describing both the generation rate of first generation of neurtral pions and the pair creation rate of UHE-photons, the distribution of UHE-photons with distance$l$, assuming rectilinear propagation, is,
%In this case the flux of photons at distance $l$ from the source is approximately 
\begin{eqnarray}
\frac{N_{\gamma}(E_{\gamma},l)}{N_p(E_{p},0)}\approx \frac{l_{\gamma\gamma}(e^{-l/l_{p\gamma}}-e^{-l/l_{\gamma\gamma}})}{(l_{p\gamma}-l_{\gamma\gamma})}
\end{eqnarray}
for which $l_{p\gamma}$ is the interaction length for an UHE-proton of
energy $E_{p}$, $l_{\gamma\gamma}$ the interaction length for an
UHE-photon of energy $E_{\gamma}$, the two energies being related
through $E_{\gamma}\approx 0.5K_pE_{p}$, where $K_p$ is the
inelasticity of the $p\gamma$ collision \cite{Stecker:1968}. 
In this expression $N_\gamma(E_{\gamma},l)$ is the number of photons of
energy $E_{\gamma}$ at distance $l$ from the source, similarly
$N_{p}(E_{p},0)$ is the number of protons of energy $E_{p}$
injected at the source.
%The factor of 0.5 follows from each $\pi^{0}$ decaying into two photons. 

\begin{figure}[h!]
\begin{center}
{\includegraphics[angle=-90,width=0.97\linewidth]{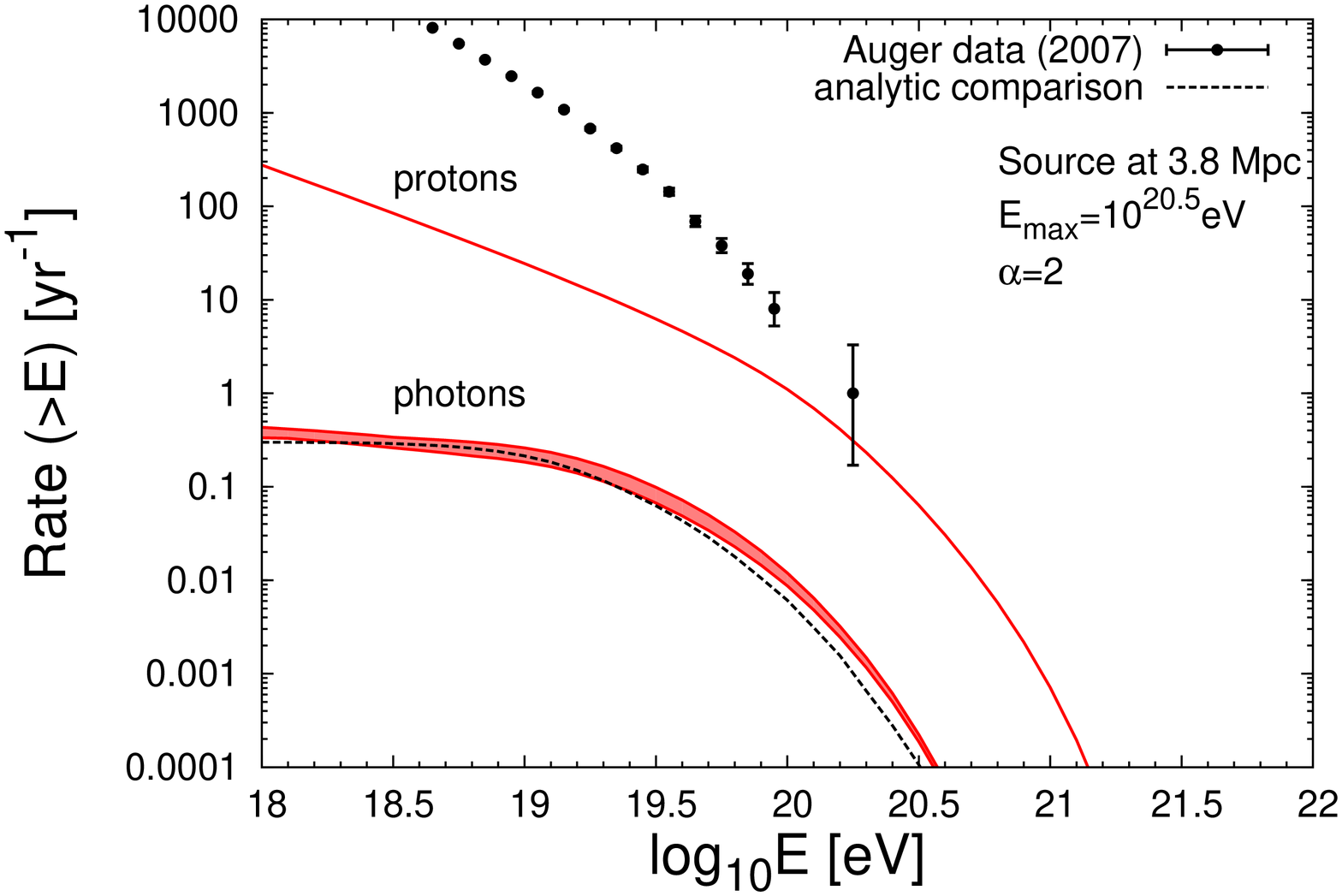}}
{\includegraphics[angle=-90,width=0.97\linewidth]{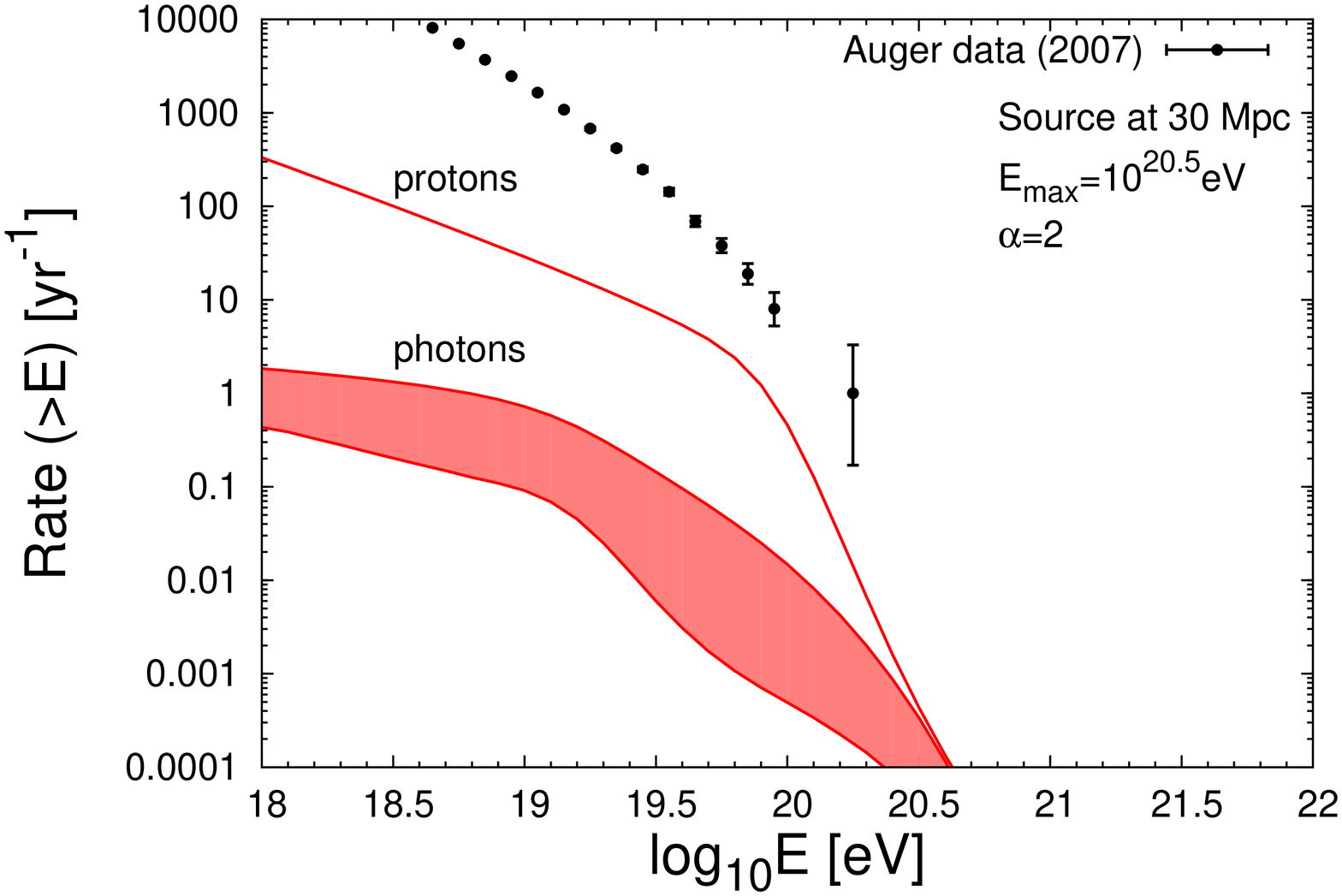}}
\caption{The expected UHE photon arrival rate from a single source 3.8~Mpc (upper plot)
and 30~Mpc (lower plot) away. For these plots, the arriving UHECR flux has been
normalised such that 10\% of the PAO events \cite{Abraham:2008ru} above $6 \times 10^{19}$~eV
came from the source. An injection index $\alpha=2$ and a cut-off energy $E_{\rm max}=10^{20.5}$~eV
were assumed.  The uncertainty in the UHE photon flux is represented by the width of the 
band drawn for the photon flux.  
For the very nearby source, 3.8~Mpc away, the analytic result for the photon flux is shown for comparison.
These results bound the source distance range for which a local UHECR source leads to a
strongly enhanced UHECR photon fraction, as demonstrated in Fig.~9 of \cite{Taylor:2008jz}}
\label{proton_photon_flux}
\end{center}
\end{figure}

In order to more accurately describe proton propagation as well as explore 
the range of possibilities for the propagation of electromagnetic energy 
through intergalactic space we here employ a 
Monte-Carlo (MC) method for UHECR proton and photon/electron propagation 
as described in \cite{Taylor:2008jz}.
In this MC UHE photon pair-production interactions with background photons 
as well as UHE electron/positron production through charged pion decay are considered, 
both giving rise to a flux of secondary UHE-electrons. The environment through which
these electrons propagate determines the ratio of the energy that these
electrons input into a new generation of UHE photons via inverse Compton
(IC) scattering on CRB photons to that lost
in much lower energy synchrotron emission through interactions with the 
IGMFs. Both the strength of the IGMFs in the relevant 
volume and the density of the CRB are very uncertain. For our calculations, intergalactic
magnetic fields in the range $10^{-9}-10^{-12.5}$~G are considered,
(B-fields weaker than $10^{-12.5}$~G play no significant role for
UHE-electron cooling on 30~Mpc distance scales).
% POSSIBLE ADDITION
However, we note that the filling factor of IGMFs in galaxy cluster regions carry a
large degree of uncertainty. We here assume that the sub $\mu$G central field \cite{Kim:1989} 
drops off sufficiently quickly with distance that UHE-electron propagation occurs 
predominantly outside such regions. 
We consider radio backgrounds ranging from the 
radio component of the CMB alone up to the measured, and possibly foreground contaminated,
value of \cite{Clark:1970}. For the range of radio background values we
consider, the uncertainty in the interaction length of $10^{19}$~eV
photons is $\approx$25\%, increasing by a factor of $\sim$2 for
photons with energy $3\times 10^{19}$~eV.
% is 2-2.5 Mpc (@ 10^19 eV) and 3-6.7 Mpc (@ 10^19.5 eV)
In the results that follow we account for the uncertainty in both the
CRB and IGMFs, by showing the complete range of photon fluxes obtained from 
the ranges of both of these quantities.

Fig.~\ref{proton_photon_flux} shows the expected integrated rate (per
year) of UHECR protons and photons from a source 3.8~Mpc away and 
30~Mpc away arriving to the (complete) PAO. A power law injection spectrum 
at source with index $\alpha=$2 and an exponential cut-off at
$10^{20.5}$~eV was assumed. The flux of UHECRs from the source has been normalised
such that it accounts for 10\% (a doublet) of events above $6 \times 10^{19}$~eV 
recorded by the PAO in the data set \cite{Abraham:2008ru} 
(corresponding to just over 2 events in the data which had an exposure of 7000~km$^{2}$~steradian~years). 
%(corresponding to 2.4 events per year for an annual exposure of 8162~km$^{2}$~steradian). 
Note that the \emph{luminosity} required by an UHECR source normalised in this way
varies with declination, due to the non-uniform sky coverage of the PAO. 
In the first panel of Fig.~\ref{proton_photon_flux} we show that this method gives 
a very similar result to the analytical description given above.

\begin{figure}[t]
\begin{center}
{\includegraphics[angle=-90,width=0.97\linewidth]{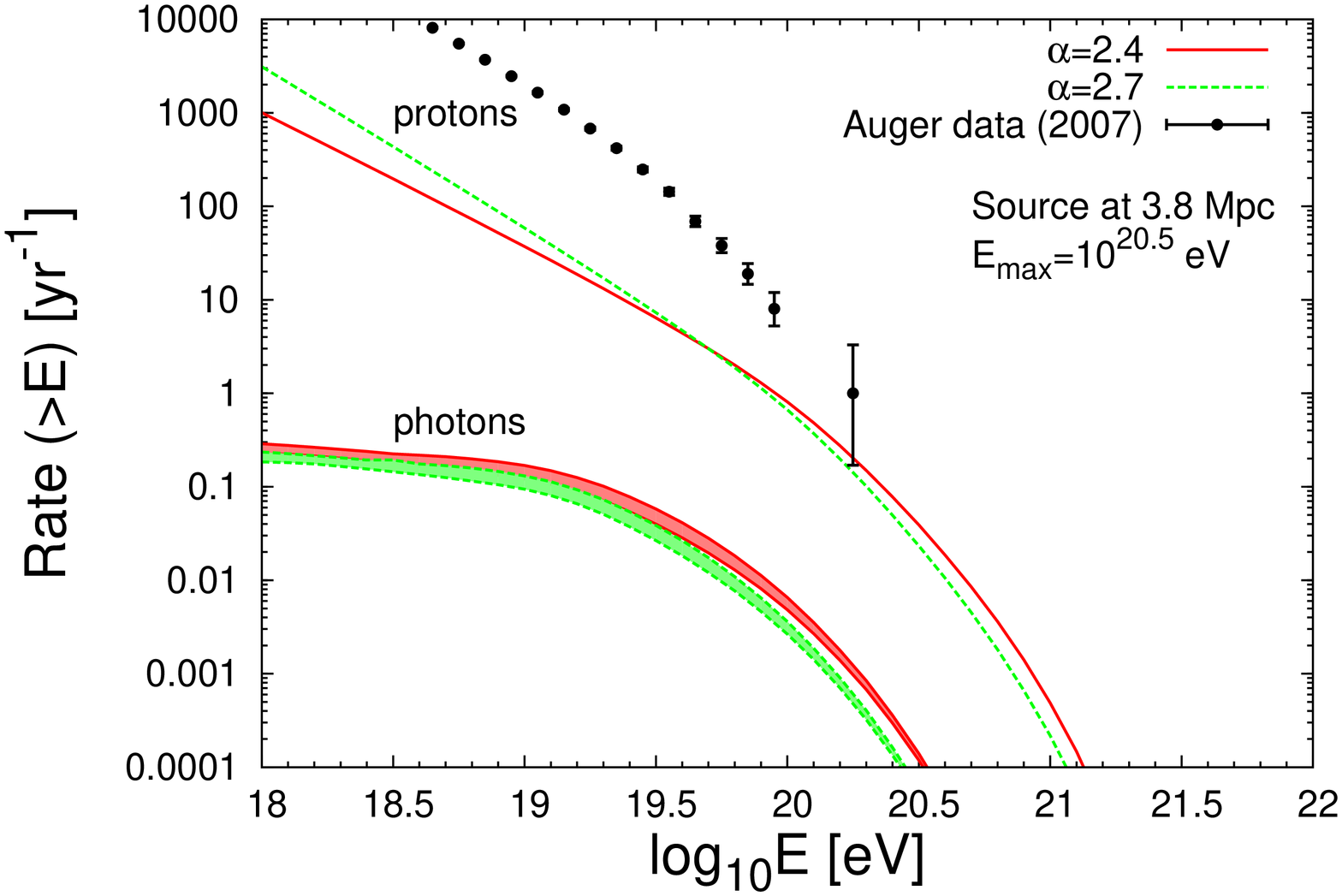}}
{\includegraphics[angle=-90,width=0.97\linewidth]{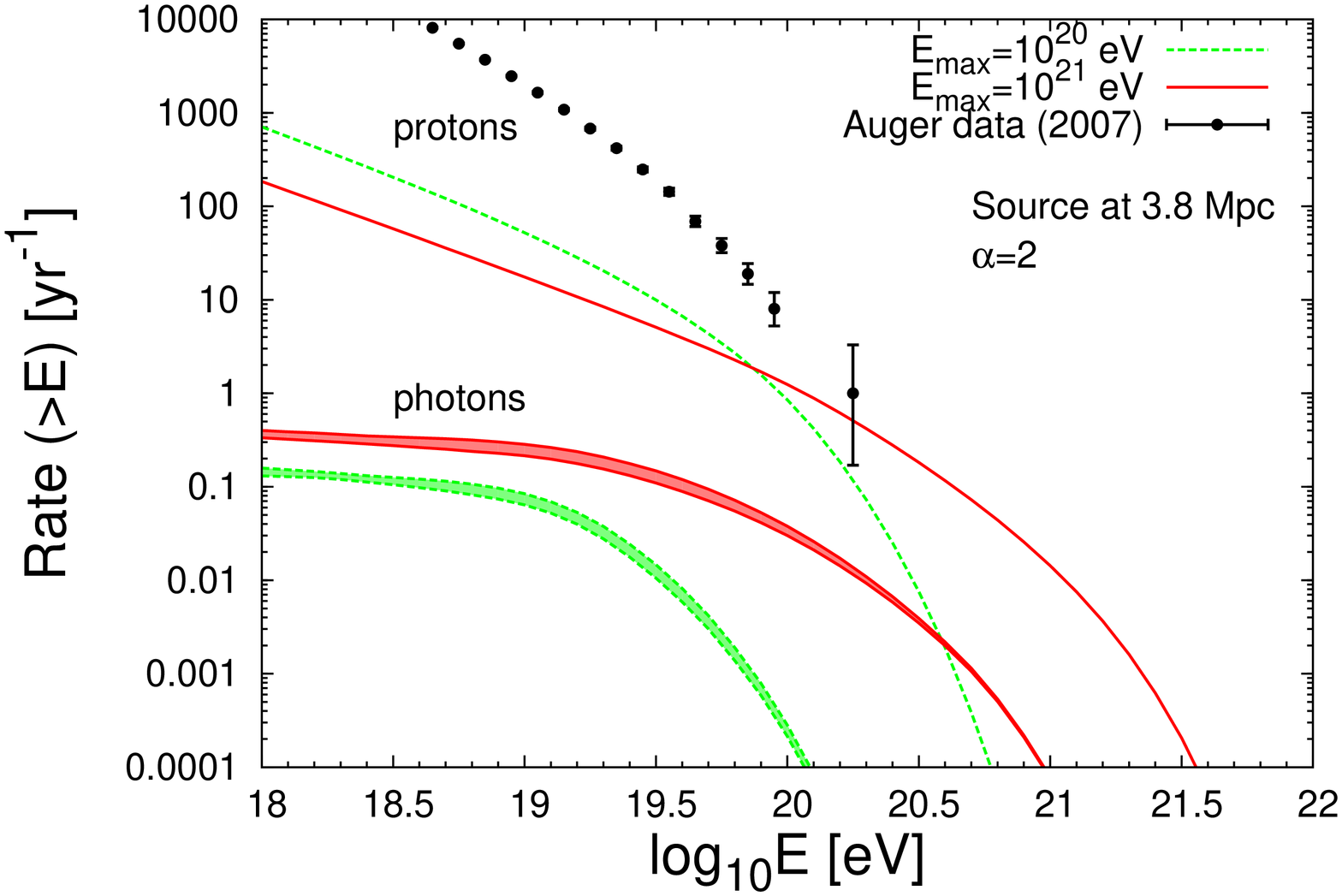}}
\caption{The expected protons and photon arrival rates as for
Fig.~\ref{proton_photon_flux}, from a source 3.8~Mpc away for
which both the injection cut-off energy ($E_{\rm max}$) and injection 
index ($\alpha$) have been varied.}
\label{proton_photon_flux2}
\end{center}
\end{figure}

Fig.~\ref{proton_photon_flux2} shows the impact on the expected
arriving photon rate from a source 3.8~Mpc away for different assumed source spectra. 
In the upper panel different injection spectral indices of $\alpha=2.4$ and
$\alpha=2.7$ are shown for comparison. In the case of a steep
injection spectrum produced by the source, with $\alpha=2.7$, a large
contribution to the arriving UHECR flux at lower
energies is expected. Such a large contribution from a nearby source ($\approx 10\%$) 
over a large energy range would likely also lead to a
detectable degree of anisotropy in the UHECRs with energies $<6\times 10^{19}$~eV. Furthermore
such a scenario would place uncomfortable requirements on the required source luminosity
unless the spectrum becomes harder (flatter) at lower energies.
As expected, the effect of a different injection spectrum on the
photon rate is weak since the rate of UHECR protons have here been
normalised to the integral flux at energies just above the photo-pion threshold
and hence UHE photon production. The effect of
reducing the maximum acceleration energy in the source, however,
results in a significant reduction in the photon flux, as can be seen
in the lower plot of Fig.~\ref{proton_photon_flux2}.

Accompanying this flux of UHE photons arriving to the Galaxy, a secondary 
flux of UHE electrons is also generated. This is similar in magnitude to the UHE photon flux, 
though it cuts off at energies somewhat below the equivalent cut-off in the UHE photon flux.  
These UHE electrons have similar (within a factor of 2) interaction lengths compared to those of
the UHE photons of the same energy as they propagate through the intergalactic medium between Cen A
and our Galaxy. Being charged particles these electrons will undergo deflections in
the IGMFs. For $\sim 10^{-10}$~G IGMFs, as suggested by \cite{Dolag:2004kp}, 
$\sim 0.5^{\circ}$ size deflections of $10^{19}$~eV electrons can be expected for propagation 
distances of $\sim$Mpc (assuming the field is coherent on this scale).

Once the UHE electrons have entered the magnetised halo of our Galaxy, they can
be expected to lose energy through synchrotron emission in a time 
$\tau_{loss} \sim 1.3~B_{\mu}^{-2}E_{19}^{-2}$~years, where $B_{\mu}$ is the magnetic field 
in $\mu$G and $E_{19}$ is the electron energy in units of $10^{19}$~eV. 
All of the electron's energy is lost in a small fraction, within $0.002^{\circ}$, %$\sim 6\times 10^{-6}B_{\mu}^{-1}E_{19}^{-2}$ 
of its gyration period, rapidly converting its energy into $\sim 1.5~B_{\mu} E_{19}^{2}$~TeV photons. 
It is therefore possible that the detection of UHE photons from a nearby source 
may be accompanied by a corresponding TeV gamma-ray flux originating from the same direction.
The angular extension of this emission most likely being dominated by the angular spread of the
arriving UHE electrons, expected to be $\sim 0.5^{\circ}$ for $10^{-10}$~G IGMFs.
This $\gamma$-ray signal may be detectable by next generation Cherenkov telescope arrays 
such as CTA \cite{Hermann:2007ve}.

From figure~\ref{proton_photon_flux}, the expected rate of 
$>10^{19}$~eV photons from a source 3.8~Mpc away, such as Cen A, is 
seen to be $0.2-0.3$ events/year in Auger 
assuming it is responsible for 10\% of the $>6\times 10^{19}$~eV flux. 
Should a large fraction of events in the current PAO dataset
originate from Cen A, this implied rate will of course increase accordingly. 
However, the discrimination of photons from protons does not have
100\% efficiency. If the data from fluorescence detectors is used 
the efficiency is reduced by an order of magnitude due to their $\sim$10\% duty
cycle.
Surface detectors have a 100\% duty-cycle
but the overall photon efficiency (for a background free sample) in current analyses is close
to 25\% \cite{Aglietta:2007yx}. No UHE photon candidate from anywhere on the
sky has so far been identified. In the search for these photons from an
individual source their selection criteria could be
significantly relaxed - the background from misidentified protons
should be less than $\sim$1\% in an 0.1 steradian region around the
source - rather than in the $\pi$ steradian field-of-view of Auger. A
photon signal from single nearby sources is therefore potentially
detectable using the PAO with a few years of data.

Given the small statistics involved it is unlikely that the
\emph{non-detection} of UHE photons from an object like Cen A would 
place strong constraints on UHECR origin and composition in the near future. 
However, the detection of one or more $>10^{19}$~eV photons from a nearby object,
particularly in combination with $\sim10^{20}$~eV UHECR correlated to it,
would provide extremely strong evidence for that object being an UHECR proton 
source. 
Along with this, the angular separation of such spatially correlated UHE photon 
and UHECR events could provide constraints on the intervening Galactic and IGMFs.
Furthermore, should future observations detect the associated TeV emission
from an UHE photon and UHECR source, the angular spread of this TeV emission 
may also allow the determination of the strength of intervening IGMFs.

Given the importance of UHE photons as a diagnostic tool in UHECR studies
we believe that effective discrimination between photons and nuclei
should be a major design criterion for future UHECR detectors such as
JEM-EUSO \cite{EUSO} and Auger North \cite{Nitz:2007ur}.

\begin{acknowledgments}
The authors are part of team \#151 of the International Space Science Institute 
and gratefully acknowledge their support for this work.
We thank Felix Aharonian and Olaf Reimer for their helpful and insightful discussions.
JAH is supported by a Science and Technology Facilities Council (UK) Advanced
Fellowship.
\end{acknowledgments}


\begin{thebibliography}{99}

%GZK from HiRes
\bibitem{hiresgzk}
Abbasi {\it et al.} (HiRes Coll.) 2008, Phys. Rev. Lett., 100, 1101

% Abraham data 2008
\bibitem{Abraham:2008ru}
 J.~Abraham {\it et al.}  [Pierre Auger Collaboration],
 %``Observation of the suppression of the flux of cosmic rays above $4\times
 %10^{19}$eV,''
 Phys.\ Rev.\ Lett.\  {\bf 101}, 061101 (2008)
 [arXiv:0806.4302]% [astro-ph]].
 %%CITATION = PRLTA,101,061101;%%

% Auger Science paper
\bibitem{AugerCollaboration:2007}
 The Auger Collaboration
 Science 318, 938 (2007)

% GZK papers
\bibitem{Greisen:1966jv}
 K.~Greisen,
 %``End To The Cosmic Ray Spectrum?,''
 Phys.\ Rev.\ Lett.\  {\bf 16}, 748 (1966).
 %%CITATION = PRLTA,16,748;%%
\bibitem{Zatsepin:1966jv}
 G.~T.~Zatsepin and V.~A.~Kuzmin,
 %``Upper limit of the spectrum of cosmic rays,''
 JETP Lett.\  {\bf 4}, 78 (1966)
 [Pisma Zh.\ Eksp.\ Teor.\ Fiz.\  {\bf 4}, 114 (1966)].
 %%CITATION = ZFPRA,4,114;%%

\bibitem{Unger:2007mc}
  M.~Unger and f.~t.~P.~Collaboration,
  %``Study of the Cosmic Ray Composition above 0.4 EeV using the Longitudinal
  %Profiles of Showers observed at the Pierre Auger Observatory,''
  arXiv:0706.1495 %[astro-ph].
  %%CITATION = ARXIV:0706.1495;%%

\bibitem{Archbold:2003vw}
  G.~Archbold and P.~V.~Sokolsky  [HiRes Collaboration],
  %``UHECR composition studies with HiRes stereo data,''
%\href{http://www.slac.stanford.edu/spires/find/hep/www?irn=5903017}{SPIRES entry}
%{\it Prepared for 28th International Cosmic Ray Conferences (ICRC 2003), Tsukuba, Japan, 31 Jul - 7 Aug 2003}
{\it ICRC, Tsukuba, Japan, 31 Jul - 7 Aug 2003}

% Wolfendale on nearby Fe sources
\bibitem{Wibig:2007pf}
 T.~Wibig and A.~W.~Wolfendale,
 %``Heavy Cosmic Ray Nuclei from Extragalactic Sources above 'The Ankle',''
 arXiv:0712.3403 %[astro-ph].
 %%CITATION = ARXIV:0712.3403;%

\bibitem{blasicas}
  C. Ferrigno, P. Blasi and D. De Marco, Astropart. Phys. {\bf 23} (2005) 211

\bibitem{gabici}
  S. Gabici and F. Aharonian, Phys. Rev. Lett. {\bf 95} (2005) 1102

\bibitem{Aharonian:1992qf}
  F.~A.~Aharonian, P.~Bhattacharjee and D.~N.~Schramm,
  %``Photon / proton ratio as a diagnostic tool for topological defects  as the
  %sources of extremely high-energy cosmic rays,''
  Phys.\ Rev.\  D {\bf 46} (1992) 4188.
  %%CITATION = PHRVA,D46,4188;%%

% possible paper for the photon fraction (up-to-date)
\bibitem{Collaboration:2009qb}
 T.~P.~A.~Collaboration,
 %``Upper limit on the cosmic-ray photon fraction at EeV energies from the
 %Pierre Auger Observatory,''
 arXiv:0903.1127 %[astro-ph.HE].
 %%CITATION = ARXIV:0903.1127;%%

% paper on photon fraction + nearest distance of source (investigates most effects)- NEEDS ADDING TO PAPER
\bibitem{Gelmini:2005wu}
 G.~Gelmini, O.~Kalashev and D.~V.~Semikoz,
 %``GZK Photons as Ultra High Energy Cosmic Rays,''
 J.\ Exp.\ Theor.\ Phys.\  {\bf 106} (2008) 1061
 [arXiv:astro-ph/0506128].
 %%CITATION = JTPHE,106,1061;%%

% Taylor paper on proton and photon prop. in nearby Universe
\bibitem{Taylor:2008jz}
 A.~M.~Taylor and F.~A.~Aharonian,
 %``The Spectral Shape and Photon Fraction as Signatures of the GZK-Cutoff,''
 arXiv:0811.0396 %[astro-ph].
 %%CITATION = ARXIV:0811.0396;%%

% Moskalenko CenA etc
\bibitem{Moskalenko:2008iz}
 I.~V.~Moskalenko, L.~Stawarz, T.~A.~Porter and C.~C.~Cheung,
 %``On the Possible Association of Ultra High Energy Cosmic Rays with Nearby
 %Active Galaxies,''
 arXiv:0805.1260 %[astro-ph].
 %%CITATION = ARXIV:0805.1260;%%

% Distance to Cen A
\bibitem{Rejkuba:2003xt}
 M.~Rejkuba,
 %``The distance to the giant elliptical galaxy NGC 5128,''
 Astron.\ Astrophys.\  {\bf 413} (2004) 903
 [arXiv:astro-ph/0310639].
 %%CITATION = AAEJA,413,90

% Cen A power
\bibitem{Evans:2004ev}
%  D.~A.~Evans, R.~P.~Kraft, D.~M.~Worrall, M.~J.~Hardcastle, C.~Jones, W.~R.~Forman and S.~S.~Murray,
  D.~A.~Evans {\it et al.},
  %``Chandra and XMM-Newton Observations of the Nucleus of Centaurus A,''
  Astrophys.\ J.\  {\bf 612} (2004) 786
  [arXiv:astro-ph/0405539].
  %%CITATION = ASJOA,612,786;%%

% Cen A luminosity (TeV)
\bibitem{Aharonian:2009xn}
  :.~F.~Aharonian  [HESS Collaboration],
  %``Discovery of very high energy gamma-ray emission from Centaurus A with
  %H.E.S.S,''
  arXiv:0903.1582 %[astro-ph.CO].
  %%CITATION = ARXIV:0903.1582;%%

% paper on probing Cen A with TeV photons + UHE neutrinos
\bibitem{Kachelriess:2008qx}
  M.~Kachelriess, S.~Ostapchenko and R.~Tomas,
  %``High energy radiation from Centaurus A,''
  arXiv:0805.2608 %[astro-ph].
  %%CITATION = ARXIV:0805.2608;%%

% AGN acc. site requirements paper
\bibitem{Norman:1995}
  C.~A.~Norman, D.~B.~Melrose, and A.~Achterberg
  Ap.~J. 454, 60-68 (1995)

% Cen A acc. paper (refers to it in part of it)
\bibitem{Casse:2001be}
  F.~Casse, M.~Lemoine and G.~Pelletier,
  %``Transport of cosmic rays in chaotic magnetic fields,''
  Phys.\ Rev.\  D {\bf 65} (2002) 023002
  [arXiv:astro-ph/0109223].
  %%CITATION = PHRVA,D65,023002;%%

% Andrew's recent paper- Cen A acc. in radio lobes paper
\bibitem{O'Sullivan:2009sc}
 S.~O'Sullivan, B.~Reville and A.~M.~Taylor,
 %``Stochastic particle acceleration in the lobes of giant radio galaxies,''
 arXiv:0903.1259 %[astro-ph.HE].
 %%CITATION = ARXIV:0903.1259;%%

% Stecker's paper on pgamma inelasticity
\bibitem{Stecker:1968}
 F.~W.~Stecker,
 Phys. Rev. Lett. 21, 1016 - 1018 (1968)

% B-fields in galaxy cluster regions
\bibitem{Kim:1989}
  K.~T.~Kim, P.~P.~Kronberg, G.~Giovannini and T.~Venturi,
  Nature 341, 720 - 723 (1989)

% Radio background observations
\bibitem{Clark:1970}
 T.~A.~Clark, L.~W.~Brown and J.~K.~Alexander,
 Nature 228, 847-849

% paper on local extragalactic magnetic fields
\bibitem{Dolag:2004kp}
  K.~Dolag, D.~Grasso, V.~Springel and I.~Tkachev,
  %``Constrained simulations of the magnetic field in the local universe and
  %the propagation of UHECRs,''
  JCAP {\bf 0501} (2005) 009
  [arXiv:astro-ph/0410419].
  %%CITATION = JCAPA,0501,009;%%

% CTA
\bibitem{Hermann:2007ve}
 G.~Hermann, W.~Hofmann, T.~Schweizer and M.~Teshima  [CTA Collaboration],
 %``Cherenkov Telescope Array: The next-generation ground-based gamma-ray
 %observatory,''
 arXiv:0709.2048 %[astro-ph].
 %%CITATION = ARXIV:0709.2048;%%

%Auger SD photon Limit
\bibitem{Aglietta:2007yx}
  J.~Abraham {\it et al.}  [Pierre Auger Collaboration],
  %``Upper limit on the cosmic-ray photon flux above 10$^{19}$ eV using the
  %surface detector of the Pierre Auger Observatory,''
  Astropart.\ Phys.\  {\bf 29} (2008) 243
  [arXiv:0712.1147 [astro-ph]].
  %%CITATION = APHYE,29,243;%%

%JEM-EUSO
\bibitem{EUSO}
  M.~Teshima,
  Astronomische Nachrichten, {\bf 328} (2007) 7

% paper for Auger North
\bibitem{Nitz:2007ur}
 D.~Nitz  [Pierre Auger Collaboration],
 %``The Northern Site of the Pierre Auger Observatory,''
 arXiv:0706.3940 %[astro-ph].
 %%CITATION = ARXIV:0706.3940;%%

\end{thebibliography}
\end{document}